\documentclass{mem}
\usepackage{natbib}\usepackage{txfonts}\usepackage{balance}
\usepackage{graphicx}
\usepackage[a4paper]{hyperref}
\idline{76}{1}
\begin{document}

\title{
Using $\delta$~Cep Stars to study Northern Dwarf Irregular Galaxies of
the Local Group
}

   \subtitle{}

\author{
C.A. \,G{\"o}ssl\inst{1} 
\and J. \, Snigula\inst{2,1}
\and U. \, Hopp\inst{1,2}
          }

\offprints{C.A. G{\"o}ssl}
 
\institute{
Universit{\"a}tssternwarte M{\"u}nchen,
Scheinerstr. 1,
D-81679 M{\"u}nchen, Germany
\and
Max-Planck-Institut f{\"u}r extraterrestrische Physik,
Giessenbachstr.,
D-85748 Garching, Germany\\
\email{cag@usm.uni-muenchen.de}
}

\authorrunning{C.A. \,G{\"o}ssl}

\titlerunning{$\delta$~Cep Stars in Northern Dwarf Irregular Galaxies}

\abstract{
Dwarf galaxies in the Local Group provide a unique astrophysical
laboratory.
Despite their proximity some of these systems still lack a reliable
distance determination as well as studies of their stellar content and
star formation history.
We present first results of our survey of variable stars in a sample
of six Local Group dwarf irregular galaxies.
We describe observational strategies and data reduction, and discuss the
lightcurves of newly found and rediscovered $\delta$~Cep stars in
DDO~216, Leo~A and GR8.
Based on these data, we present newly derived independent Cepheid
distances.
Other variable stars found in our survey are discussed in a related
article of this volume (Snigula et al.).
}
\maketitle{}

\section{Introduction}
The main aim of the Wendelstein (WST) monitoring project is to determine
numbers and properties of the bright variable stars in six northern
dwarf irregular galaxies:
LGS~3 (Pisces),
UGCA~92 (EGB~0427+63),
DDO~69 (Leo~A),
DDO~155 (GR8),
DDO~210 (Aquarius), and
DDO~216 (Pegasus).
Those will be used
to put further constraints on their stellar content,
and thus on their evolutionary history,
\citep[][Snigula et al. this volume]{2004ASPC..310...70S}
and for distance estimates.
Here we present the results of our survey for classical Cepheids in
Leo~A, DDO~216, and GR8.
We derive distance moduli and compare them with recently published ones.
\section{Observations}
\begin{table}[t]
  \caption{Epochs per filter, telescope and object.
    \label{o25_epochs}}
  \begin{center}
    \begin{tabular}{lrrrrr}
      \hline
      & \multicolumn{2}{c}{WST} & \multicolumn{3}{c}{CA} \\
      object & $B$ & $R$ & $B$ & $R$ & $I$\\
      \hline
      Leo~A & 31 & 94 & 13 & 47 & 29 \\
      DDO~216 & 17 & 70 & -- & 2 & -- \\
      GR8 & 23 & 106 & 14 & 49 & 31 \\
      \hline
    \end{tabular}
  \end{center}
\end{table}
The relatively small 0.8~m Wendelstein telescope has the necessary
long-term availability for monitoring projects (i.e. 130 clear nights
per year, unrestricted access).
Although the telescope does not take full advantage of the good
($\ll 1"$) seeing quality of the site, we regularly obtain images of 1
to $1.5"$ FWHM \citep{2001A&A...379..362R}.
The Wendelstein observations, starting with test observations in 1999,
sparsely sample a five year interval in $R$ and $B$ filter bands.
We added observations in the $R$, $B$ and $I$-bands, obtained
with the 1.23~m telescope at Calar Alto (CA) observatory
(Tab.~\ref{o25_epochs}).
The Wendelstein data are used to find variable sources and to
determine their periodicities, while the CA data, if present, serve as
an independent consistency check.
The typical limiting magnitude of an individual exposure is about
$R\sim22.5$
($M_R=-2.0$ for Leo~A, $M_R=-2.5$ for DDO~216, $M_R=-4.0$ for GR8,
respectively).
Thus, we have access to all kinds of red long period variable stars,
blue and red irregular variables,
and also Novae and Supernovae but so far none of either type of
exploding stars were detected.
RR~Lyr stars are certainly too faint, while classical
$\delta$~Cep stars are well within the limits of our data.
Tab.~\ref{o25_epochs} displays the number of collected and reduced
epochs so far.
We restrict the results presented in this article to periods of
$P<130$ days, the longest period values known for $\delta$~Cep stars.
\begin{table*}
  \caption{\footnotesize
    Parameters of our $\delta$~Cep stars:
    identifier,
    position,
    most significant Lomb period, significance ($p$-level),
    flux averaged apparent $R$-band magnitude,
    RMS error of $R$-band magnitude.
    Error of period $\delta P<0.01$d.
    V01 to V05 = {\bf $\bullet$} and
    V06 to V07 = {\bf $\blacktriangle$} in Fig.~\ref{o25_leo_plr}.
    P01 to P03 = {\bf $\bullet$} and
    P04 to P06 = {\bf $\blacktriangle$} in Fig.~\ref{o25_pegasus_plr}.
    \label{o25_data_tab}}
\begin{center}
\begin{tabular}{cccccccc}
\hline
Id & RA-2000 & Dec-2000 & period & significance & $<R_{\bar{f}}>$ &
$\delta<R_{\bar{f}}>$ \\
 & [h] & [deg] & [d] & [$p$] & [mag] & [mag] \\
\hline
V01 & 09:59:28.679 & +30:44:35.38 & 6.487 & 6.93e-10 & 20.62 & 0.11 \\
V02 & 09:59:27.762 & +30:44:57.42 & 1.685 & 8.78e-03 & 21.45 & 0.25 \\
V03 & 09:59:23.914 & +30:45:13.06 & 3.354 & 5.86e-04 & 21.47 & 0.25 \\
V04 & 09:59:29.115 & +30:43:48.70 & 2.049 & 9.80e-03 & 22.10 & 0.44 \\
V05 & 09:59:30.472 & +30:44:03.65 & 1.685 & 3.43e-05 & 22.26 & 0.52 \\
V06 & 09:59:25.918 & +30:44:36.70 & 1.607 & 2.61e-04 & 21.68 & 0.30 \\
V06$a$ & & & 2.630 & 1.63e-04 & & \\
V07 & 09:59:25.672 & +30:44:41.79 & 1.564 & 5.47e-04 & 22.07 & 0.44 \\
\hline
P01 & 23:28:32.702 & +14:45:15.49 & 3.889 & 1.31e-02 & 21.03 & 0.24 \\
P02 & 23:28:36.217 & +14:44:02.64 & 3.712 & 1.06e-02 & 21.71 & 0.45 \\
P03 & 23:28:37.310 & +14:43:45.33 & 2.642 & 5.07e-03 & 22.23 & 0.73 \\
P04 & 23:28:33.753 & +14:44:27.68 & 3.188 & 2.93e-02 & 21.09 & 0.26 \\
P05 & 23:28:28.990 & +14:44:41.88 & 1.118 & 1.75e-02 & 21.41 & 0.34 \\
P06 & 23:28:32.340 & +14:44:47.06 & 8.593 & 5.99e-02 & 22.17 & 0.69 \\
\hline
GR8 & 12:58:41.389 & +14:13:09.47 & 15.436 & 9.47e-07 & 21.47 & 0.07\\
\hline
\end{tabular}
\end{center}
\end{table*}
\begin{figure*}[]
\begin{center}
\resizebox{0.95\hsize}{!}{\includegraphics{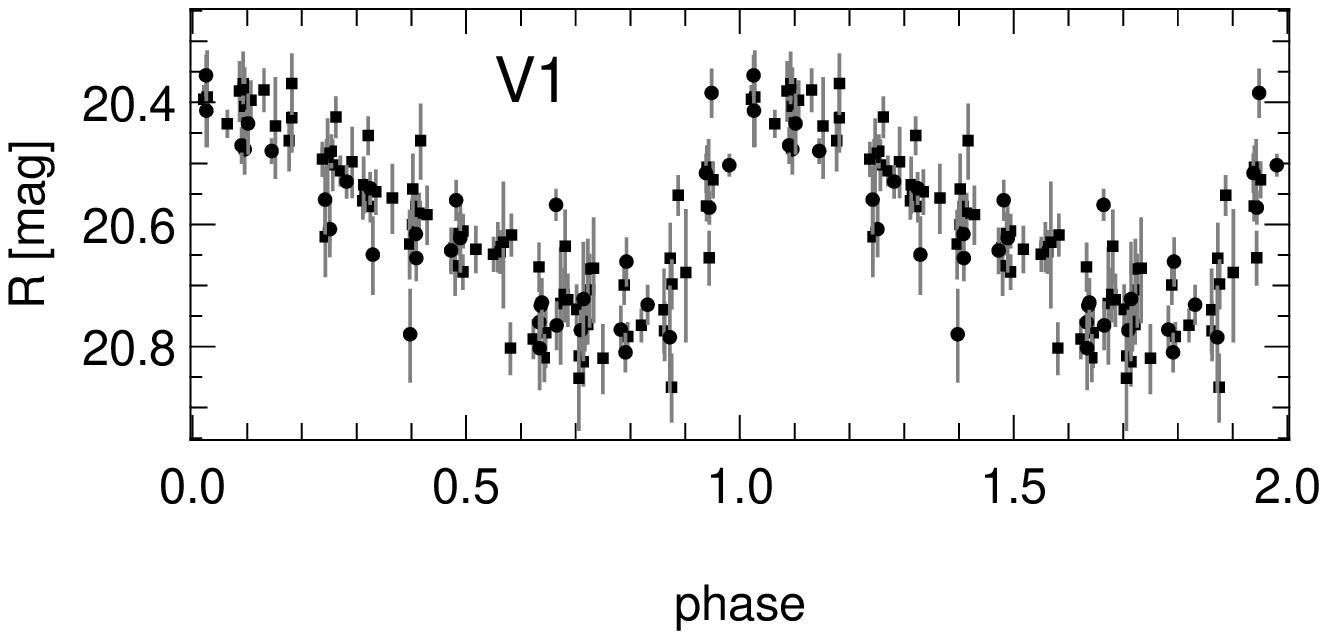}
\hspace*{5mm}\includegraphics{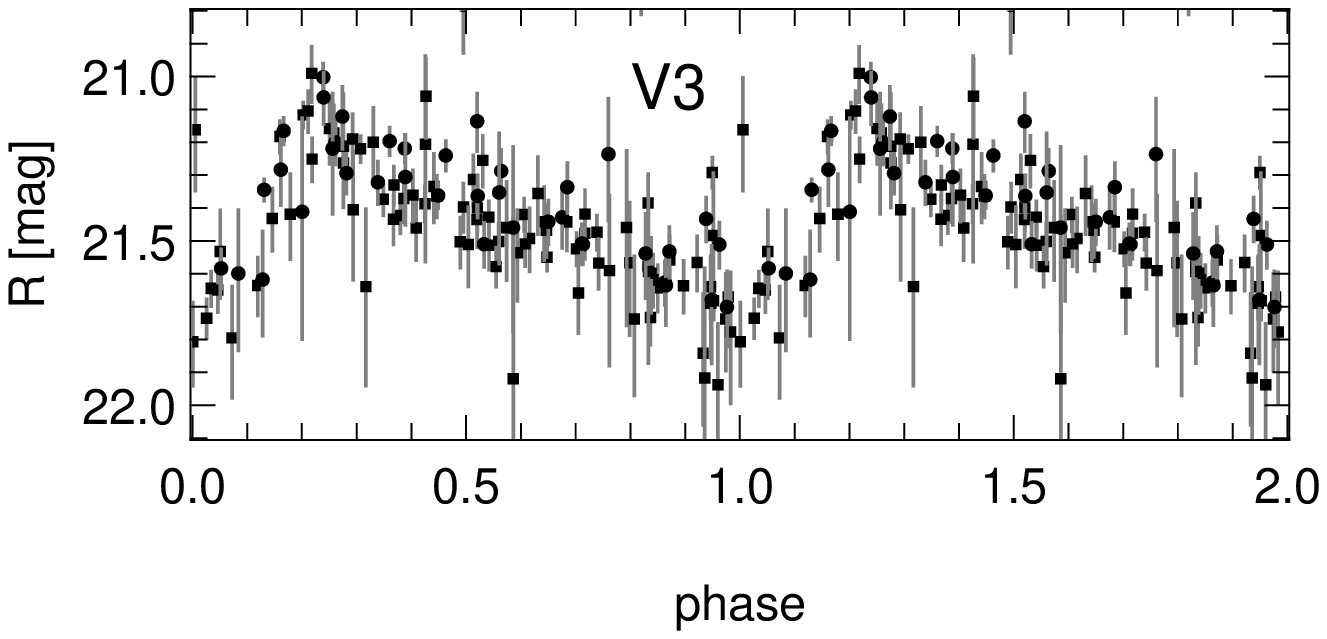}
\hspace*{5mm}\includegraphics{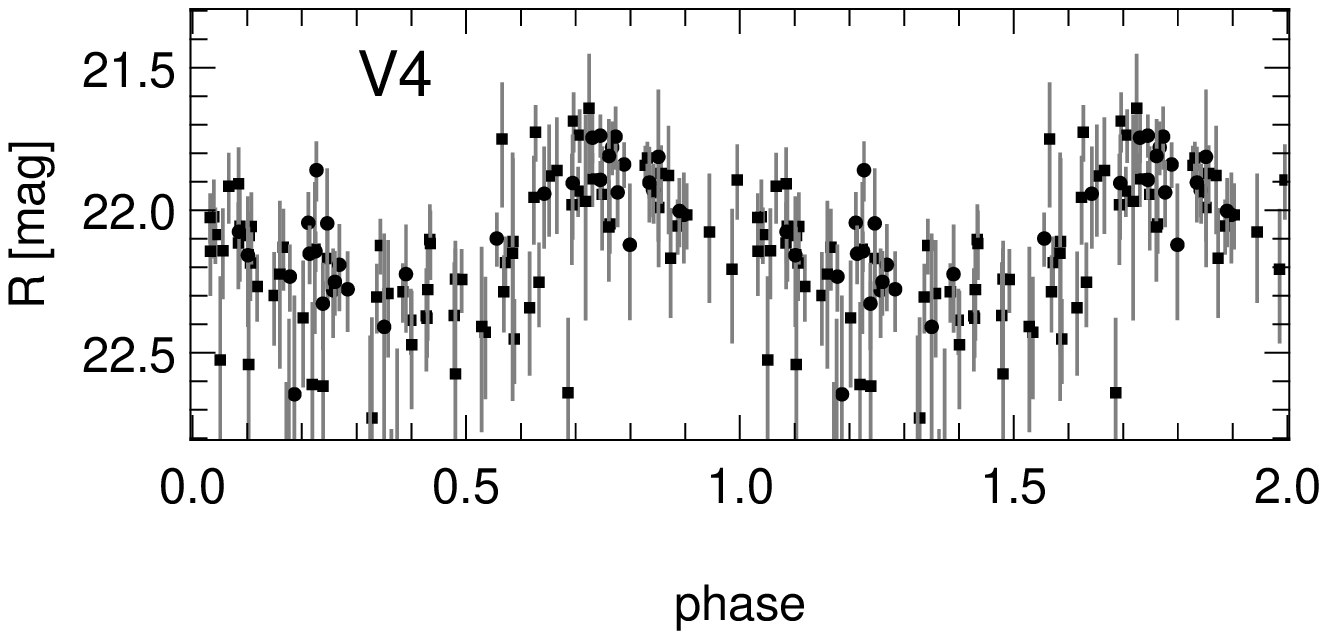}}\\
\vspace*{2mm}
\resizebox{0.95\hsize}{!}{\includegraphics{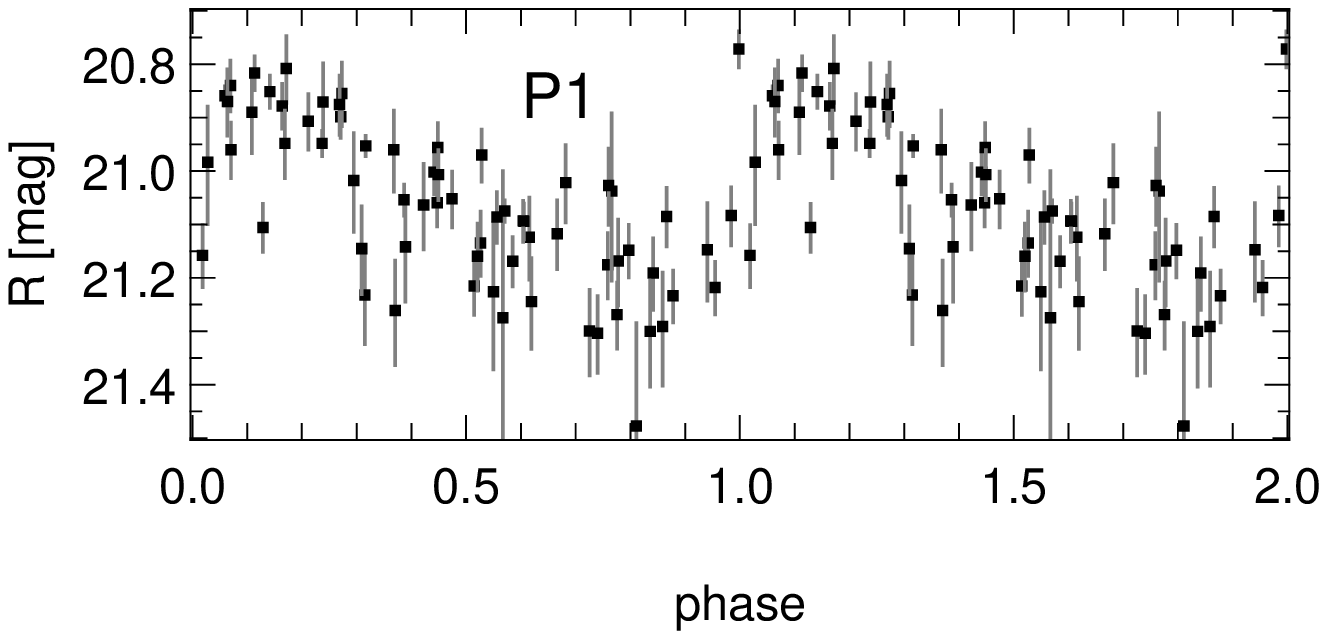}
\hspace*{5mm}\includegraphics{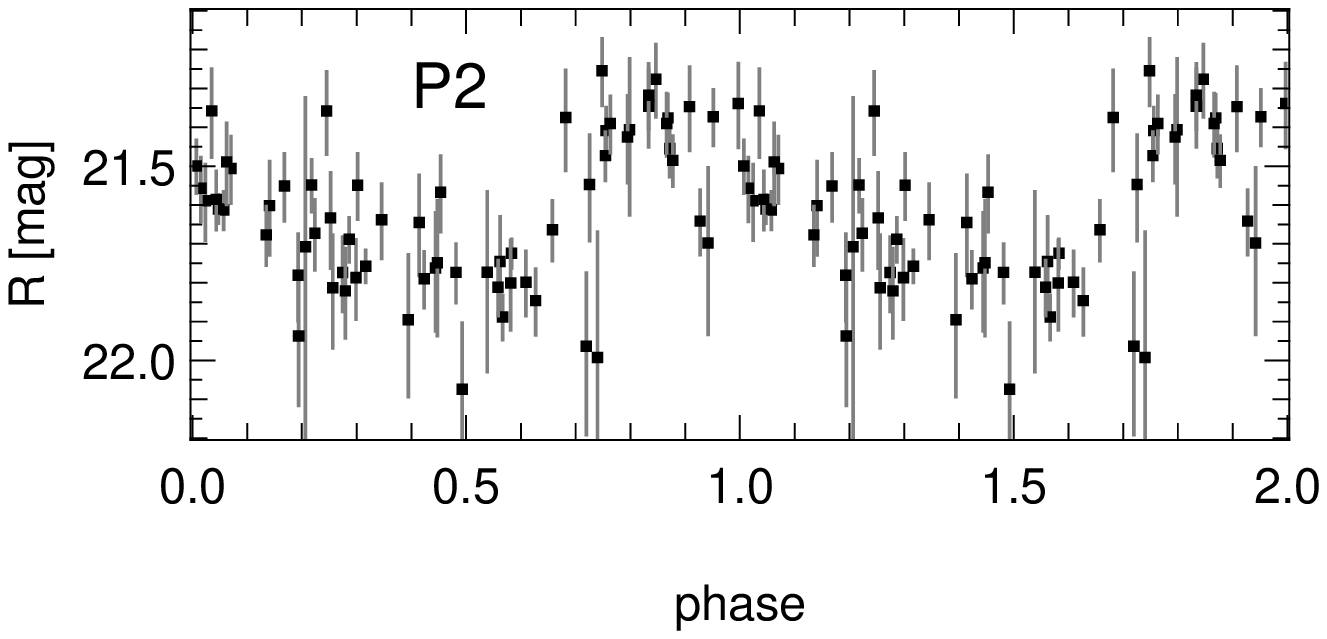}
\hspace*{5mm}\includegraphics{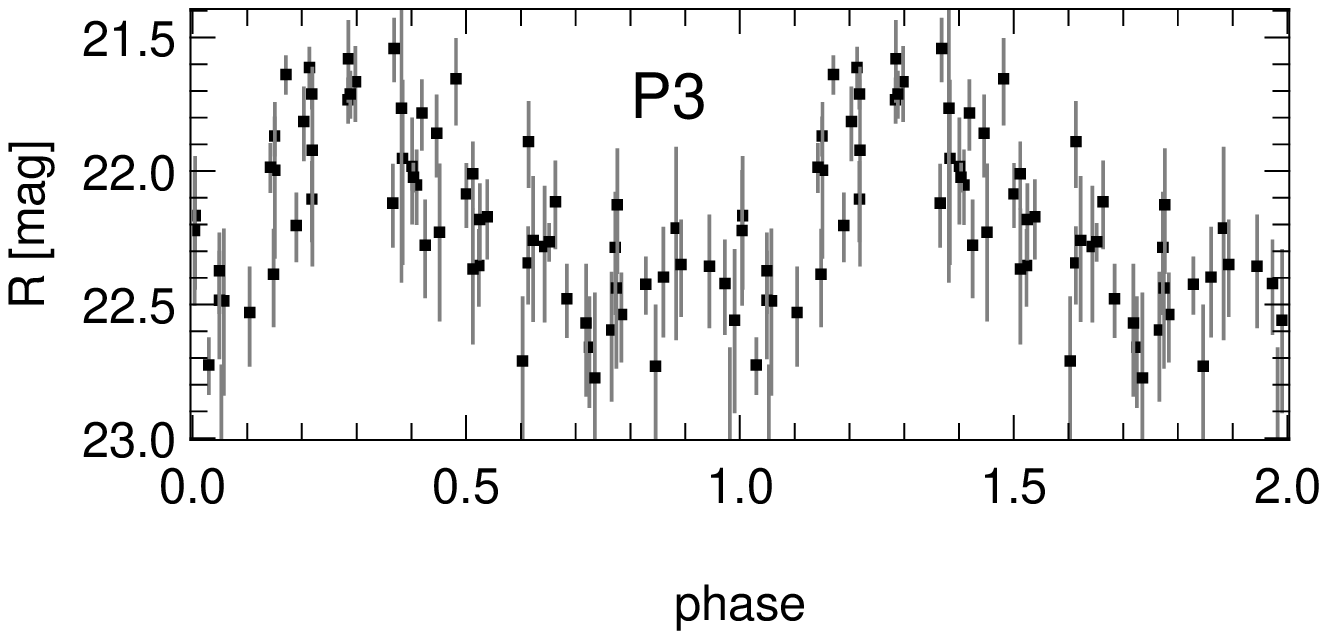}}
\end{center}
\caption{
\footnotesize
Phase convolved $R$-band lightcurves for the $\delta$~Cep
stars used to derive the distance moduli.
$\blacksquare$ = Wendelstein 0.8~m, $\bullet$ = Calar Alto 1.23~m,
$+$ = measurements having $S/N < 1$ obtained at either site,
all plotted with $1 \sigma$ error bars.
Top: Leo~A, bottom: DDO~216.
\label{o25_lcs}
}
\end{figure*}
\section{Data Reduction}
All images were bias subtracted, flat-fielded, and cleaned of particle
events.
After an astrometrical alignment signal-to-noise maximising stacks per
night were built.
For every stack, representing an epoch, a difference image against a
common deep reference frame was created applying an implementation
\citep{2002A&A...381.1095G,2003adass..12..229G}
of the Alard algorithm
\citep{1998ApJ...503..325A}.
These difference images were finally convolved with a stellar PSF.
Our codes propagate individual pixel errors through every step of the
data reduction.

We find variable star candidates by first building a (cumulative) mask
frame counting where and how often individual difference frames
deviate from zero by at least $1 \sigma$ (i.e. propagated error).
For all candidates indicated by this variability mask, a
\citet{1976Ap&SS..39..447L} algorithm in the interpretation from
\citet{1982ApJ...263..835S} is used to search for periodic signals.

\begin{figure}[]
\begin{center}
\resizebox{0.9\hsize}{!}{\includegraphics[clip=true]{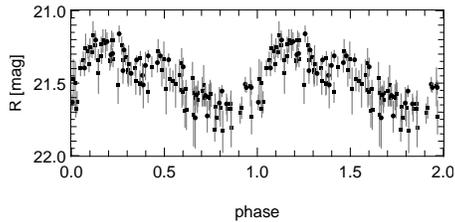}}
\end{center}
\caption{
\footnotesize
Phase convolved $R$-band lightcurves for the $\delta$~Cep
star in GR8. Symbols see caption Fig~\ref{o25_lcs}.
}
\label{o25_gr8_lcs}
\end{figure}

To get rid of false and problematic classifications we apply rigorous
selection criteria.
\citep[See detailed discussion in][]{2005A&A..LeoA..Hopp}.
Finally we fit the $R$-band period luminosity relation (PLR) for
fundamental mode (FM) LMC Cepheids
$M_R = -3.04 (\log P - 1.0) - 4.48 [\pm 0.25]$
of \citet{1991PASP..103..933M}
corrected for galactic extinction following
\citet{1998ApJ...500..525S} to the remaining candidate(s) of each
object.
Tab.~\ref{o25_data_tab} lists the found $\delta$~Cep stars,
Fig.~\ref{o25_lcs} and \ref{o25_gr8_lcs} show the lightcurves of all
stars used for the distance moduli.
\section{Results}
\begin{figure}[]
\begin{center}
\resizebox{0.9\hsize}{!}{\includegraphics[clip=true]{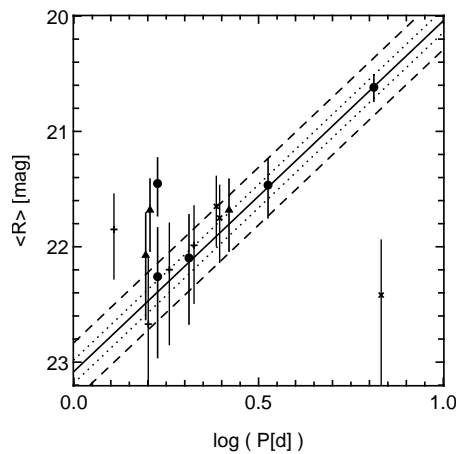}}
\end{center}
\caption{
\footnotesize
PLR for Leo~A.
Solid line:
Best fitting R-band PLR for FM LMC Cepheids
of V01, V03, and V04
(Tab.~\ref{o25_data_tab}),
extinction corrected by $A_R=0.055$:
$24.47\pm0.10\pm 0.06_{\mathrm{ZP}}$.
V02 is disregarded as being brightened by crowding, V05 for being
beyond the scope of the applied PLR.
Dotted lines: Propagated error of the fit.
Dashed lines: Error of the PLR.
{\bf $\bullet \blacktriangle$}: $\delta$~Cep stars of
Tab.~\ref{o25_data_tab}.
{\bf $+ \times$}: Add. $\delta$~Cep candidates described in
\citet{2005A&A..LeoA..Hopp}.
}
\label{o25_leo_plr}
\end{figure}
\begin{figure}[]
\begin{center}
\resizebox{0.9\hsize}{!}{\includegraphics[clip=true]{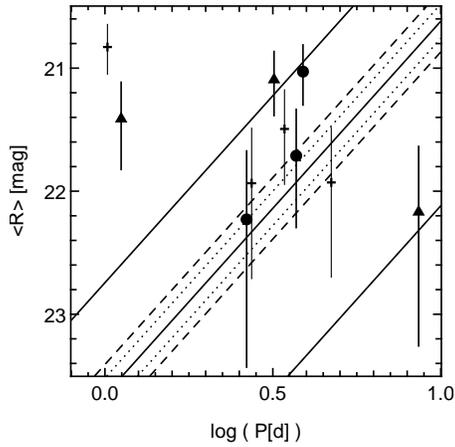}}
\end{center}
\caption{
\footnotesize
PLR for DDO~216.
Central solid line:
The R-band PLR for FM LMC Cepheids
extinction corrected by $A_R=0.176$.
Upper solid line:
$1^{\mathrm{st}}$ overtone PLR assuming a -0.92 mag offset
to the fundamental mode relation.
Lower solid line:
Type II PLR assuming a +1.5 mag offset to the fundamental
mode relation of classical $\delta$~Cep stars.
Best-fitting PLR of P01, P02, and P03:
$24.92\pm0.20\pm 0.06_{\mathrm{ZP}}$.
Dotted lines: Propagated error of the fit.
Dashed lines: Error of the PLR.
{\bf $\bullet \blacktriangle$}: $\delta$~Cep stars of
Tab.~\ref{o25_data_tab}.
{\bf $+$}: Add. $\delta$~Cep candidates matching most, but not all
of our selection criteria.
}
\label{o25_pegasus_plr}
\end{figure}
\subsection{Leo\,A}
For Leo~A we find a distance modulus of
$m-M=24.47\pm0.10\pm0.06_{\mathrm{ZP}}$ (Fig.~\ref{o25_leo_plr})
which is consistent with the findings of 
\citet{2002AJ....123.3154D},
\citet{1998AJ....116.1244T}, and \citet{2002AJ....124..896S}.
\citeauthor{2002AJ....123.3154D} searched for RR~Lyr
stars in Leo~A and derived $m-M=24.51\pm0.12$.
\citeauthor{1998AJ....116.1244T} used a combination of ground-based
and HST data to derive a tip-of-the-red-giant-branch (TRGB) distance
of $m-M=24.5\pm0.2$.
They also used red clump stars yielding a discrepant
$m-M=24.2\pm0.2$.
\citeauthor{2002AJ....124..896S}'s even deeper HST data allowed the
detection of red horizontal branch stars and concluded that they and
the TRGB are in better agreement with $m-M=24.5\pm0.2$.
\subsection{DDO\,216}
Fitting the \citeauthor{1991PASP..103..933M} FM PLR to P02, P03, and
to P01 (the latter with a doubled period, see below) we derive a
distance modulus of
$m-M=24.92\pm0.20\pm 0.06_{\mathrm{ZP}}$ (Fig.~\ref{o25_pegasus_plr})
for the Pegasus dwarf irregular.
P01 to P04 are four out of six candidates already independently
proposed by \citet{1994ApJ...437L..27A}.
He could not provide any period solutions lacking a sufficient number
of observed epochs.
Using HST imaging data \citep{1998AJ....115.1869G} we find P04 to be
artificially brightened by crowding while P01 seems to be pulsating in
the $1^{\mathrm{st}}$ overtone mode.
We confirm the TRGB distance $m-M=24.9\pm 0.1$ of
\citeauthor{1994ApJ...437L..27A}.
\subsection{GR\,8}
In GR8 we only find one $\delta$~Cep star with
$m_R=21.47\pm0.07$ and $P=15.44$d (Fig.~\ref{o25_gr8_lcs}).
We derive a distance modulus of
$m-M=26.45\pm0.07\pm0.06_{\mathrm{ZP}}(\pm0.25_{\mathrm{PLR}})$
taking into account a galactic extinction of $A_R=0.07$
\citep{1998ApJ...500..525S}.
Since this distance is based on a single detection the full intrinsic
scatter of the PLR has to be considered as an additional uncertainty.
This $\delta$~Cep star had already been discovered by
\citet{1995AJ....109..579T} with $m_{<r>}=22.12$ and $P=16.166$d.
They derived a distance modulus of $m-M=26.75\pm 0.35$
which \citet{1998AJ....116.1227D} claim to be consistent with their
HST based TRGB observations.
\section{Summary}
We confirmed by independent discoveries and measurements of
$\delta$~Cep stars recently derived distance moduli of the three
Northern Local Group dwarf irregular galaxies Leo~A, DDO~216, and
GR8.
\bibliographystyle{aa}
\end{document}